\documentclass[12pt]{article}
\usepackage[export]{adjustbox}
\usepackage{graphicx}
\usepackage{graphics}
\usepackage{epsfig}
\usepackage{amssymb}
\usepackage{amsmath}
\usepackage{soul,xcolor}
\usepackage{color}
\usepackage{textcomp}
\usepackage{latexsym}
\usepackage{physics}
\usepackage[square,comma,sort&compress]{}
\usepackage[noadjust]{cite}

\usepackage[margin=2.5cm,footskip=0.6cm]{geometry}
\begin{document}

\begin{center}
{\Large{\bf Kinetics of Self-Organization in One Component Active Matter Systems}} \\
\ \\
by \\
Avinash Chauhan and Prasenjit Das$^*$\footnote{prasenjit.das@iisermohali.ac.in} \\
Department of Physical Sciences, Indian Institute of Science Education and Research -- Mohali, Knowledge City, Sector 81, SAS Nagar 140306, Punjab, India. \\
\end{center}

\begin{abstract}
\noindent We study an inherently nonequilibrium system consisting of self-propelled particles using molecular dynamics simulation in $d=2$. The system self-organizes through quorum-sensing interactions characterized by an interaction range $\lambda$ and a parameter $c_{\mathrm{th}}$. When the local concentration sensed by a particle, represented by an auxiliary variable $c_i(t)$ exceeds $c_{\mathrm{th}}$, its self-propulsion velocity abruptly drops to zero, leading to the emergence of particle rich domains. For small $\lambda$, we observe a droplet-like morphology of clusters, which crosses over to a bicontinuous structure as $\lambda$ increases. We confirm this transition by computing the eccentricity of the domains. Next, we characterize the evolving domain morphologies using equal-time correlation function $C(r,t)$  and structure factor $S(k,t)$ of the density field. For a fixed $\lambda$, the scaled $C(r,t)$'s at different times are indistinguishable from each other, indicating dynamical scaling. However, at a fixed time $t$, the scaling functions differ for different values of $\lambda$, confirming a $\lambda$-dependent morphological transition. In the large-wavevector limit ($k\rightarrow\infty$), the tail of $S(k,t)$ consistently follows the Porod's law: $S(k,t)\sim k^{-3}$. Furthermore, the average domain size $L(t)$ exhibits a power-law growth, $L(t)\sim t^{1/3}$, independent of $\lambda$.
\end{abstract}

\newpage
\section{Introduction}\label{sec1}
Self-organization in active matter systems has emerged as a rapidly growing field of research over the last two decades \cite{Czirok1996, Neufeld1996, TONER2005, JULICHER2007, Ramaswamy2010, Lauersdorf2025}. The constituents of an active system are self-propelled agents that can consume energy from their surroundings or convert internal energy into directed motion \cite{Ramaswamy2010, Marchetti2013, OByrne2022,SMPD2025}. Active matter encompasses many examples in nature, ranging from biological tissues, bird flocks \cite{Cavagna2014, Chen2019}, fish shoals \cite{Bonner1998, Jiang2023}, animal herds \cite{Hueschen2023}, and pedestrian crowds \cite{Corbetta2023}, to synthetic systems such as microrobots \cite{Volpe2011}, Janus particles \cite{Jiang2010}, and vibrated granular matter \cite{Kudrolli2008}. The typical size of these constituents varies from a few millimeters to several centimeters. All of these systems are intrinsically out of equilibrium, which leads to several emergent phenomena, including nonequilibrium pattern formation, giant number fluctuations \cite{Toner2019}, and spontaneous flows generated by internal stresses \cite{Whitfield2017}. Such phenomena do not exist in passive systems.

In this paper, we focus on the kinetics of self-organization in active Brownian particle~(ABP) systems. The existing literature primarily attributes such phenomena to motility-induced phase separation~(MIPS) and non-reciprocal interactions~\cite{Duan2023,ivlev15,ivlev18,MJS24,saha2020,you2020}. Our study primarily focuses on self-organization driven by MIPS. In the classical MIPS, an active particle transiently reduces its speed due to interactions with its high-density surroundings, leading to an effective accumulation of particles in denser regions. This accumulation enhances the density fluctuations and further reduces the particle speed, thereby establishing a positive feedback loop that drives aggregation of particles \cite{Cates2015, Partridge2019,ATCFJS2024}. 

Starting from the seminal particle-based simulations of Vicsek \textit{et al.}~\cite{vicsek95}, many theoretical~\cite{solon2015,zkd23} and experimental efforts have been made over the past few decades to understand MIPS in ABP systems~\cite{ISA2022}. Theoretically, various particle-based and coarse-grained models have been proposed \cite{Tailleur2008}. In the context of particle-based simulations, Redner \textit{et al.} studied MIPS in a system of self-propelled Brownian spheres that interact purely through exclude volume, with no alignment interaction~\cite{RHB13}. The system phase separates into dense and dilute phases whose densities depend solely on the P\'eclet number of the spheres. Agudo-Canalejo and Golestanian studied MIPS in mixtures of chemically interacting particles~\cite{ACG19}. Chemical reactions among the species break the action–reaction symmetry of the interactions, leading to separation into phases with distinct densities and stoichiometry. Stenhammar \textit{et al.} studied the phase-separation kinetics of MIPS in $d=3$ and showed that the average domain size, $L(t)$, follows the classical Lifshitz--Slyozov~(LS) growth law, $L(t)\sim t^{1/3}$ \cite{Sten2014}. In addition, they demonstrated that the structure factor obeys Porod's law in the large wavenumber regime. 

The first coarse-grained model for MIPS was proposed by Tailleur and Cates \cite{Tailleur2008}. They derived an evolution equation for the density field by coarse-graining the single-particle dynamics. They showed that the system undergoes phase separation into dilute and dense phases when the self-propulsion speed $v$ decreases with the local particle density. Later, Curatolo  \textit{et al.} extended the study for $n$-species systems~\cite{AIC2020}. Recently,  Mondal and Das studied the domain kinetics for the model with $n=2$. They showed that $L(t)$ follows the LS law, independent of the relative activities of the species or their compositions~\cite{SMPD2024}. At the phenomenological level, Wittkowski \textit{et al.} introduced the active model B~(AMB) by adding a gradient-squared term of the order parameter, i.e., $\alpha|\vec{\nabla}\phi|^2$, to the chemical potential of passive model B~\cite{witt2014}. This additional term violates the detailed balance condition, a defining signature of nonequilibrium active matter systems. By tuning $\alpha$, they controlled the activity of the system. They observed that $L(t)$ exhibits power-law growth, $L(t)\sim t^\phi$, with an exponent $\phi$ that depends on $\alpha$. Nevertheless, they predicted the $L(t)\sim t^{1/3}$ growth law irrespective of the value of $\alpha$. Recently, Pattanayak \textit{et al.}~\cite{SSS21,SSS2} provided a detailed study of domain growth kinetics for symmetric and asymmetric active binary mixtures using AMB. They reported that the domain growth exponent exhibits a crossover from $1/3$ at early times to $1/4$ at later times. Further extension of this study using AMB+, an extension of AMB, shows that the asymptotic domain growth law is $L(t)\sim t^{1/4}$~\cite{pradeepyadav}.

There exist several experimental studies on MIPS. One of the earliest experimental studies is due to Dombrowski \textit{et al.}, who showed that bacterial systems can form large-scale coherent structures in confined geometries~\cite{dom2004}. Subsequent experimental studies on synthetic active particles, such as Janus particles, demonstrated various tunable self-assembled structures by controlling the environment and confinement. Linden \textit{et al.} showed that switching on high activity in a relatively dense system of active Janus colloids leads to rapid clustering, followed by cluster aggregation toward full phase separation~\cite{linden2019}. Theurkauff \textit{et al.} studied dynamic clustering in active colloidal suspensions with chemical signaling and showed that, in addition to solid-like and gas-like phases observed at high and low densities, respectively, a novel cluster phase emerges~\cite{it2012}. Experimental observations of the $L(t)\sim t^{1/3}$ growth law in MIPS are limited. Aforementioned studies exhibit dynamical clustering or arrested coarsening rather than the unrestricted diffusive coarsening observed in model systems, owing to hydrodynamic effects, alignment interactions, boundaries, or finite-size effects.

Our study here is motivated by the work of B\"auerle \textit{et al.}, who experimentally and numerically studied the self-organization of light-activated active particles~\cite{BFSB18}. In their model, particles interact via a quorum-sensing interaction of range $\lambda$ and self-organize for a certain range of $\lambda$ values. However, a systematic study of the self-organization kinetics for large system sizes is still lacking. In this work, we address this issue by investigating the coarsening dynamics over a wide range of $\lambda$ and other model parameters. Our results show that $L(t)$ follows the LS law, irrespective of $\lambda$. In contrast, the domain morphology depends strongly on $\lambda$: domains are nearly circular for small $\lambda$, predominantly elliptical at intermediate values of $\lambda$, and elongated for large $\lambda$. We quantify the transition by computing eccentricity of the domains. Further, we have characterized the domain morphologies by calculating equal-time correlation functions and structure factors of the density fields.

Given this background, the paper is organized as follows. We describe the model and provide the details of simulations in Sec.~\ref{sec2}. The simulation results are discussed in Sec.~\ref{sec3}. Finally, we summarize our findings in Sec.~\ref{sec4}.

\section{Details of Modeling and Simulation}\label{sec2}
We consider a system of active particles in two dimensions. Particle $i$, with position vector $\vec r_i$, interacts with particle $j$, with position vector $\vec r_j$, via a purely repulsive Weeks–Chandler–Andersen (WCA) interaction potential $U_{\rm WCA}$, given by 
\begin{equation}
\label{seq1}
U_{\rm WCA}\left(r_{ij}\right) = 4\epsilon\left[\left(\frac{\sigma}{r_{ij}}\right)^{12}-\left(\frac{\sigma}{r_{ij}}\right)^{6}+\frac{1}{4}\right].
\end{equation}
Here, $r_{ij}=|\vec r_i - \vec r_j|$ denotes the distance between the particle pair $(i,j)$, $\sigma$ is the effective particle diameter, and $\epsilon$ is the interaction strength. The cutoff distance for $U_{\rm WCA}$ is $r_{\rm cut}=2^{1/6}\sigma$. In our simulation, we set $\epsilon=100k_{\rm B}T$ to minimize overlap between pairs of particles and ensure numerical stability, where $k_{\rm B}$ is the Boltzmann constant and $T$ denotes the room temperature.

For modeling the quorum-sensing interaction, we follow the approach proposed by B\"auerle \textit{et al}~\cite{BFSB18}. We introduce a time-dependent auxiliary variable $c_i(t)$ for the $i$th particle, defined as
\begin{equation}
\label{seq2}
c_i(t)= \tilde{c}\sum_{j\ne i} \frac{\sigma}{r_{ij}(t)}{\rm exp}\left({-\frac{r_{ij}(t)}{\lambda}}\right),
\end{equation}
where $\tilde{c}$ is a constant. The parameter $\lambda$ has the dimension of length and determines the range of the quorum-sensing interaction. The propulsion speed $v_i$ depends on $c_i(t)$: it is $v_0$ when $c_i(t)$ is less than some threshold value $c_{\rm th}$, and it becomes zero when  $c_i(t)>c_{\rm th}$. Since $c_i(t)$ encodes information about the local neighborhood of the $i$th particle, the resulting change in particle motility effectively mimics a density-dependent propulsion mechanism, giving rise to MIPS. Furthermore, we assume that the propulsion orientation $\theta_i$ (with $x$-axis), remains unaffected by the modulation of motility and it shows purely diffusive dynamics.

In the overdamped limit, the equations of motion for the particles are given by
\begin{align}
\label{seq3}
\frac{d\vec{r}_i}{dt} &= v_i\left(c_i(t)\right)\hat{n}_i-\frac{D_0}{k_BT}\vec\nabla_iU_{\rm WCA} + \vec\Lambda_i(t),\\ \
\label{seq4}
\frac{d\theta_i}{dt} &= \eta_i(t).
\end{align}
Here, $\hat{n}_i=(\cos\theta_i,\sin\theta_i)$ denotes the unit vector corresponding to the orientation angle $\theta_i$ of the $i$th particle. The stochastic term $\vec{\Lambda}_i(t)$ represents Gaussian white noise with zero mean and covariance $$\langle\Lambda_{i\alpha}(t_1)\Lambda_{j\beta}(t_2)\rangle=2D_0\delta_{ij}\delta_{\alpha\beta}\delta(t_1-t_2),$$ where $D_0$ represents the translational diffusion coefficients, $\alpha$ and $\beta$ label Cartesian components, and $i$ and $j$ denote particle indices. Similar to $\vec{\Lambda}_i(t)$, $\eta_i(t)$ also has zero mean and satisfies $\langle\eta_i(t_1)\eta_j(t_2)\rangle=2D_R\delta_{ij}\delta(t_1-t_2)$, with $D_R$ being the rotational diffusion coefficients~\cite{Ridgway2023, Fily2012}.

We simulate systems of size $L_x\times L_y=512\sigma\times 512\sigma$. The particle number density is set to $\rho=0.2$, and we choose $\sigma=1$. These parameters fix the total number of particles to $N=\rho\times512\sigma\times 512\sigma\approx52428$. The system is initialized by randomly assigning particle positions $\vec r_i$ within the simulation box, such that the separation between any pair of particles is greater than $2\sigma$. The variables $\theta_i$ are initialized by random numbers sampling from a uniform distribution over the interval $[-\pi, \pi]$. Equations (\ref{seq3}) and (\ref{seq4}) are discretized in time using the Euler discretization method to update particle positions and orientations. A time step of size $\Delta t = 0.00005$ is used. To improve computational efficiency, a cutoff distance $R_{\rm cut} = 3\lambda$ is introduced for the calculation of $c_i(t)$, and a neighbor list is implemented and updated at regular intervals. At each time step, we compute $c_i(t)$ for all particles using Eq.~\eqref{seq2} and check whether $c_i(t)$ exceeds the threshold $c_{\rm th}$. If $c_i(t)>c_{\rm th}$, we set the propulsion speed of that particle to $v_i=0$; otherwise, the particle continues to move with $v_i=v_0$ during that time step, while its propulsion orientation $\theta_i$ continues to undergo rotational diffusion. Periodic boundary conditions are employed in all spatial directions. We set $\tilde{c}=1$, $v_0=50$, $D_0=1$, and $D_R=3$. Unless otherwise stated, we use $c_{\rm th}=5.0$. We treat $c_{\rm th}$ and $\lambda$ as independent simulation parameters. All the statistical results presented below are averaged over ten independent runs.

\section{Simulation Results}\label{sec3}
\begin{figure}
\centering
\includegraphics[width=0.95\textwidth]{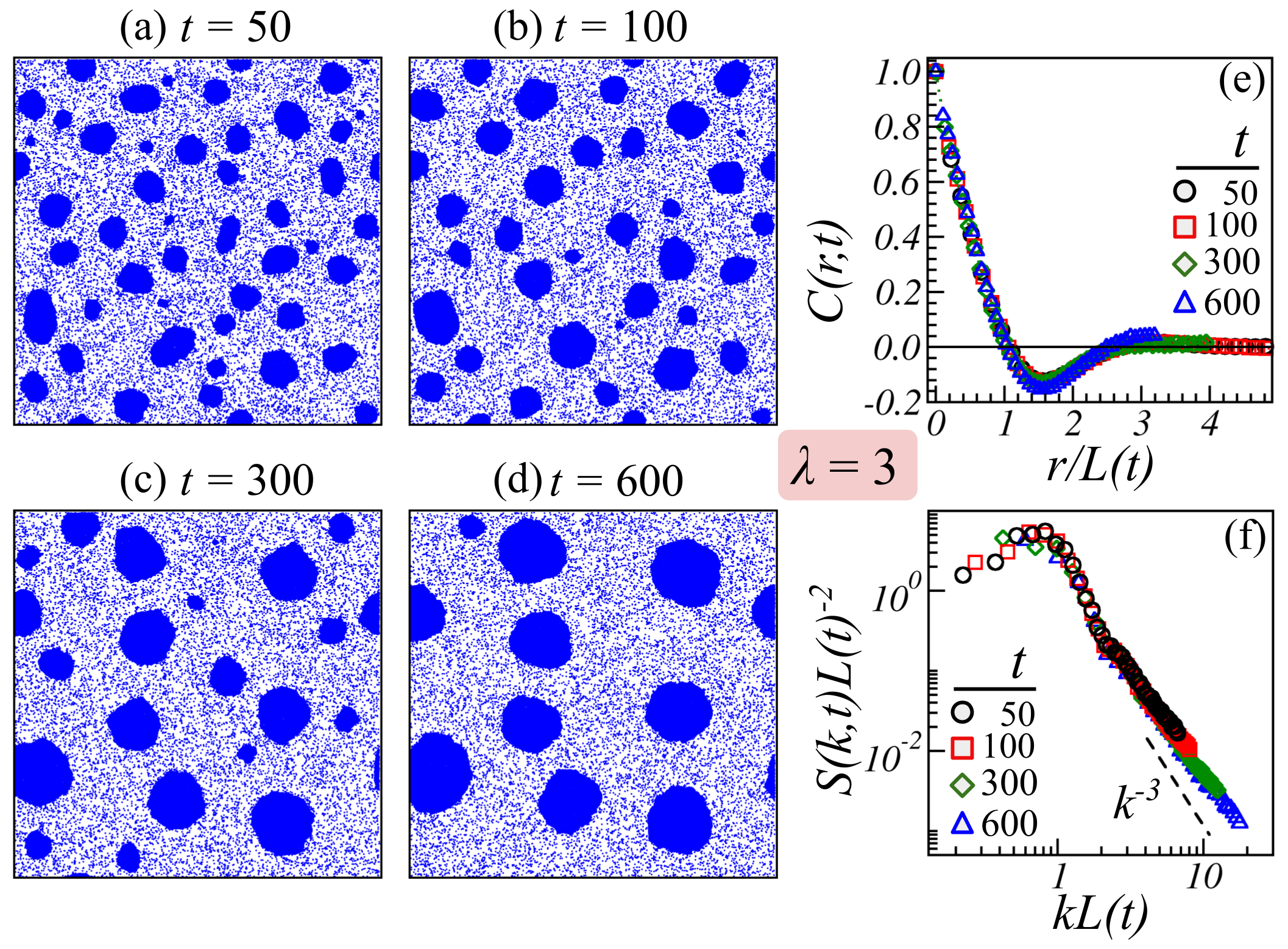}
\caption{Evolution snapshots of the system for $\lambda=3$ at different times: (a) $t=50$, (b) $t=100$, (c) $t=300$, and (d) $t=600$, as mentioned. Panel (e) shows the corresponding scaling plot of the equal-time correlation function $C(r,t)$ as a function of $r/L(t)$, where $L(t)$ is the time-dependent average domain size. Panel (f) shows the scaling plot of the structure factor $S(k,t)L(t)^{-2}$ vs. $kL(t)$ on a log-log scale. The dashed line labeled $k^{-3}$ represents Porod's law, $S(k,t)\sim k^{-(d+1)}$, for $d=2$.}
\label{Fig1}
\end{figure}

Figures~\ref{Fig1}(a)–(d) show snapshots of the evolution of particle positions at different times for $\lambda = 3$, as indicated. Clearly, the system evolves through the formation of droplet-like dense clusters or particle domains. At early times [cf. Fig.~\ref{Fig1}(a)], the number of clusters is large, while their sizes are small. As time advances [cf. Figs.~\ref{Fig1}(b)–(c)], droplets above a certain size grow, while smaller ones dissolve, as observed in the Ostwald ripening process~\cite{AO2002}. At later times, clusters with large number of particles emerge, as shown in Fig.~\ref{Fig1}(d). Typically, particles that lie well inside a cluster have zero self-propulsion speed, whereas particles near domain boundaries or in dilute regions have a $v_0$ self-propulsion speed. Also, it appears that the center of mass of the larger clusters that grow over time remains essentially stationary [cf. Figs.~\ref{Fig1}(c)–(d)].

Next, we focus on characterizing the morphology of domains by calculating the equal-time correlation function $C(r,t)$ and structure factor $S(k,t)$~\cite{SDSP2004,SPVB2009}. For the order-parameter $\phi(\vec r,t)$, $C(r,t)$ is defined as
\begin{eqnarray}
\label{seq5}
C(r,t) = \left\langle\phi(\vec{r}_1,t)\;\phi(\vec{r}_2,t)\right\rangle-\left\langle\phi(\vec{r}_1,t)\right\rangle\left\langle\phi(\vec{r}_2,t)\right\rangle.
\end{eqnarray}
Here, $r$ is the distance between two points $\vec{r}_1$ and $\vec{r}_2$, and the angular brackets represent spherical averaging over different directions and random initial conditions.
$S(k,t)$ is the Fourier transform of $C(r,t)$ and is given by
\begin{eqnarray}
\label{seq6}
S(\vec{k},t) = \int d\vec{r}\;e^{i\vec{k}.\vec{r}}\;C(\vec{r},t).
\end{eqnarray}

To define $\phi(\vec r,t)$, we map the continuum fluid configurations onto a square lattice. The system is divided into boxes of size $4\sigma \times 4\sigma$. If the number density of particles within a box exceeds the average density (here $\rho = 0.2$), we assign a value of $\phi(\vec r,t)=+1$; otherwise, we assign $\phi(\vec r,t)=-1$. This mapping renders the subsequent analysis analogous to studies of phase-separation kinetics in solid binary mixtures using Monte Carlo simulations of the spin-$1/2$ Ising model~\cite{SASKDSP10,SASKDSP12}.

Figure~\ref{Fig1}(e) shows plots of $C(r,t)$ at four different times, corresponding to the evolution snapshots shown in Figs.~\ref{Fig1}(a)–(d). By definition, $C(r=0,t)=1$, reflecting perfect self-correlation. As $r$ increases, $C(r,t)$ initially decays with distance, as the probability of finding opposite order parameters increases. At larger $r$, $C(r,t)$ becomes negative, reaching a minimum as regions with opposite order parameters dominate. Upon a further increase in $r$, $C(r,t)$ rises again owing to the mixed presence of both order-parameter states. Clearly, $C(r,t)$ at different times collapse onto a single master curve when the distance $r$ is scaled by the average domain size $L(t)$, indicating dynamical scaling. Here, the characteristic domain size, $L(t)$, is defined as the distance at which the $C(r,t)$ decays to a certain fraction of its maximum value, $C(0,t)=1$ \cite{SPVB2009, Bray1994}. Figure~\ref{Fig1}(f) shows the scaled structure factor, $S(k,t)L(t)^{-2}$, plotted as a function of $kL(t)$ at different times, as indicated. In the large-$k$ limit, all master curves approximately exhibit a $k^{-3}$ decay, consistent with \textit{Porod's law}: $S(k,t)\sim k^{-(d+1)}$ for $d=2$~\cite{porod,OP1988}. Furthermore, the tails of the scaled structure factor decrease with time (i.e., $S(k,t)L(t)^{-2}$ is larger for the black curve than for the blue curve), reflecting the growth of the average domain size.

\begin{figure}[ht!]
\centering
\includegraphics[width=0.95\textwidth]{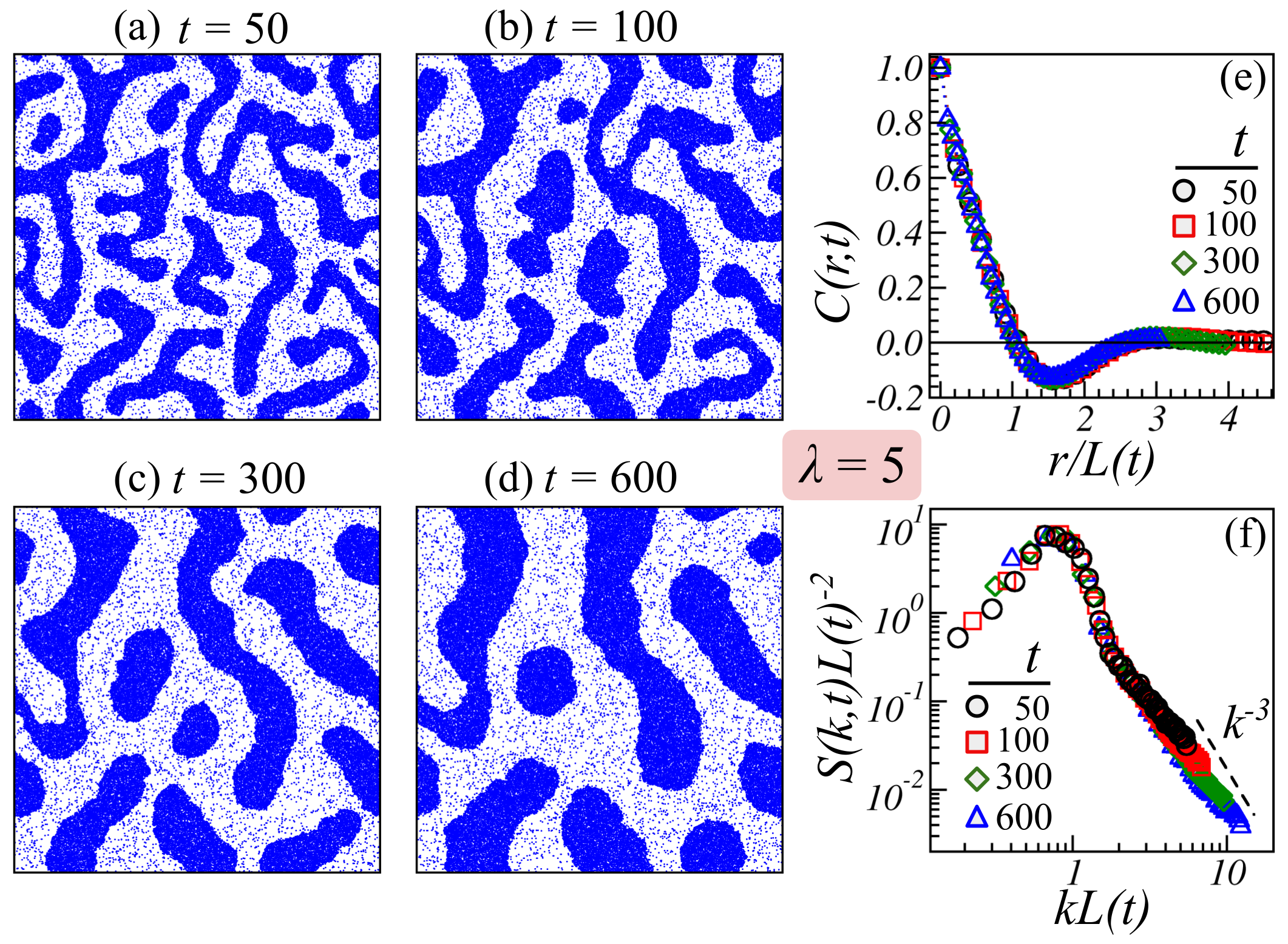}
\caption{Time evolution of the domain morphology at fixed $\lambda = 5$ is illustrated by snapshots taken at different times: (a) $t = 50$, (b) $t = 100$, (c) $t = 300$, and (d) $t = 600$. Panel (e) presents the corresponding plots of the scaled correlation function $C(r,t)$ versus $r/L(t)$, while panel (f) shows the scaling plot of $S(k,t)L(t)^{-2}$ as a function of $kL(t)$ on a log–log scale. The dashed line labeled $k^{-3}$ represents Porod’s law.}
\label{Fig2}
\end{figure}

Figures~\ref{Fig2}(a)–(d) shows snapshots of the system at different times for $\lambda = 5$, as indicated. Similar to the $\lambda = 3$ case, the system evolves by forming dense clusters of particles that grow with time. However, in sharp contrast to the $\lambda = 3$ case, we observe interconnected domain structures. We next characterize the domain morphology by computing the correlation function $C(r,t)$ and the structure factor $S(k,t)$. Similar to the $\lambda = 3$ case, the numerical data for both $C(r,t)$ and $S(k,t)$ exhibit dynamical scaling, as shown in Figs.~\ref{Fig2}(e) and \ref{Fig2}(f), respectively. Further, in the large-$k$ tail of $S(k,t)$ follows Porod’s law.

\begin{figure}
\centering
\includegraphics[width=0.95\textwidth]{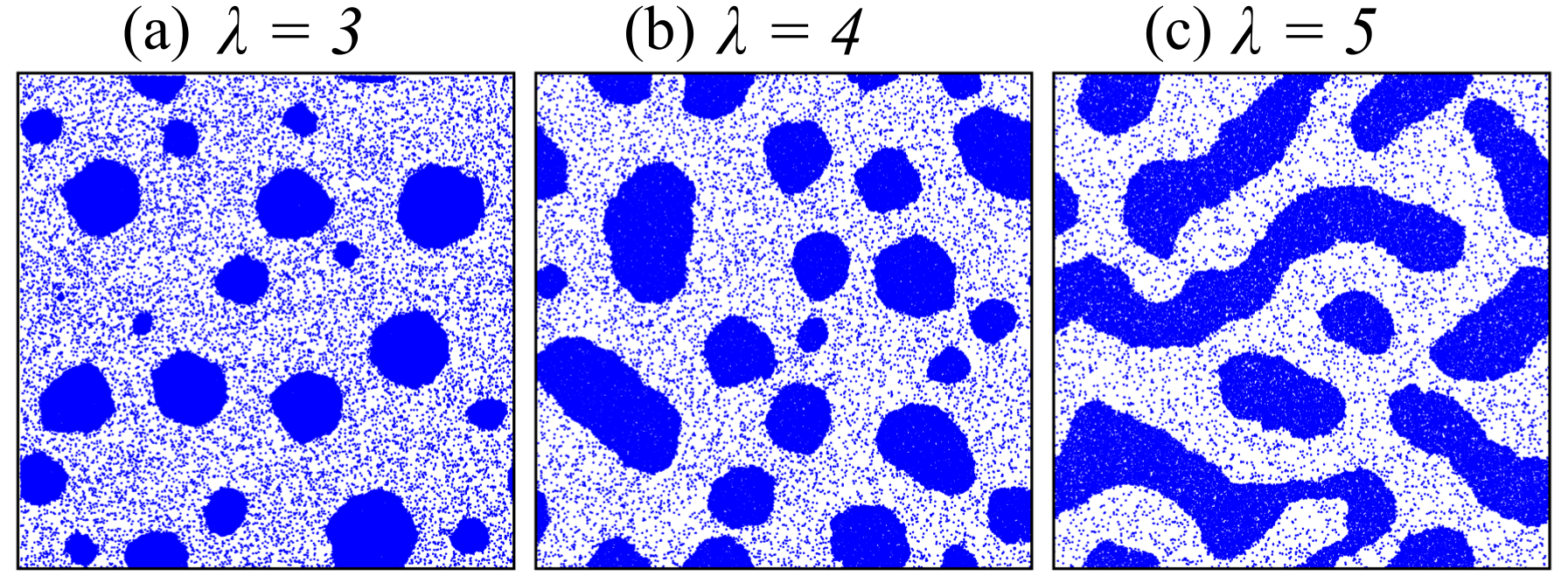}
\caption{Evolution snapshots of the system at fixed $t=300$ with different values of $\lambda$: (a) $\lambda=3$, (b) $\lambda=4$, and (c) $\lambda=5$, as mentioned. }
\label{Fig3}
\end{figure}

Next, we examine how the domain morphologies depend on $\lambda$. We compare snapshots of the system at a fixed time $t = 300$ for different values of $\lambda$, as shown in Fig.~\ref{Fig3}. For smaller $\lambda = 3$, the system exhibits a circular droplet-like morphology, as shown in Fig.~\ref{Fig3}(a). At intermediate values, for example $\lambda = 4$, the morphology consists of both circular droplets and elongated domains, as shown in Fig.~\ref{Fig3}(b). As $\lambda$ is increased further to $\lambda = 5$, the system develops an elongated and bicontinuous morphologies, as shown in Fig.~\ref{Fig3}(c). These results indicate that, for a fixed particle density and other model parameters, the domain morphology depends strongly on $\lambda$. This can be explained as follows. For small $\lambda$, the number of particles satisfying $c_i(t) > c_{\mathrm{th}}$ during the early stages of evolution is relatively small. Consequently, only a few particles have their self-propulsion velocity reduced to zero, and they are well separated from one another. These particles act as nucleation sites for domain formation, consequently the resulting domains are predominantly circular. As $\lambda$ increases, the number of nucleation sites also increases, leading to a reduction in the average separation between them. This enhances the likelihood of coalescence between neighboring circular domains, giving rise to aspherical domain structures. Upon further increasing $\lambda$, the continued increase in the number of nucleation sites and the corresponding decrease in their separation promote the formation of highly elongated domains that can span a significant fraction of the system.

\begin{figure}
\centering
\includegraphics[width=0.6\textwidth]{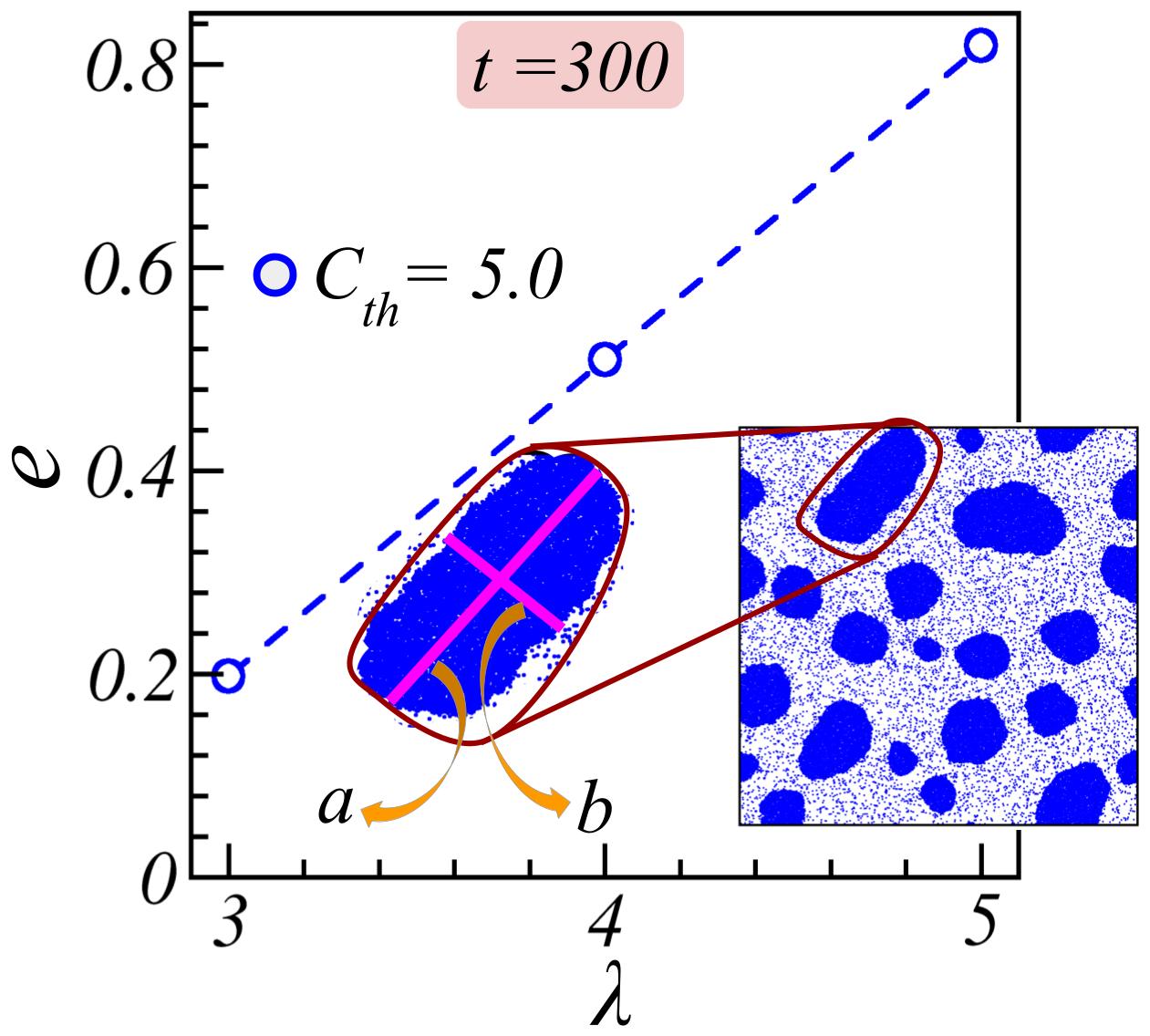}
\caption{Plot of average eccentricity $e$ of the clusters as a function of $\lambda$ at time $t = 300$, using Eq.~\eqref{seq7}. The data are obtained by averaging the eccentricities of all clusters in the system and subsequently averaging over 10 independent realizations with different initial conditions.}
\label{Fig6}
\end{figure}

To characterize the aspherical shape of the domains during evolution, we compute the eccentricity $e$ of the domains, given by
\begin{eqnarray}
\label{seq7}
e = \sqrt{1-\frac{b^2}{a^2}}.
\end{eqnarray}
Here, $a$ and $b$ denote the major and minor axes, respectively, of an aspherical shaped domain, as shown in Fig.~\ref{Fig6}. These quantities are obtained from the gyration tensor analysis. For a domain containing, $N_d$ particles, we first determine the domain centre of mass and construct the covariance (gyration) matrix from the particle position fluctuations relative to the centre of mass, as described in Ref.~\cite{Stephen2004}. The eigenvalues of the matrix, $\Gamma_1$ and $\Gamma_2$ ($\Gamma_1 \ge \Gamma_2$), correspond to the principal moments of the domain. The lengths of the major and minor axes are then given by $a=\sqrt{\Gamma_1}$ and $b=\sqrt{\Gamma_2}$, respectively. Figure~\ref{Fig6} shows the dependence of $e$ on $\lambda$, keeping other parameters fixed. Clearly, $e$ increases with $\lambda$, indicating the increasing of aspherical shape of the domains.

\begin{figure}
\centering
\includegraphics[width=1.0\textwidth]{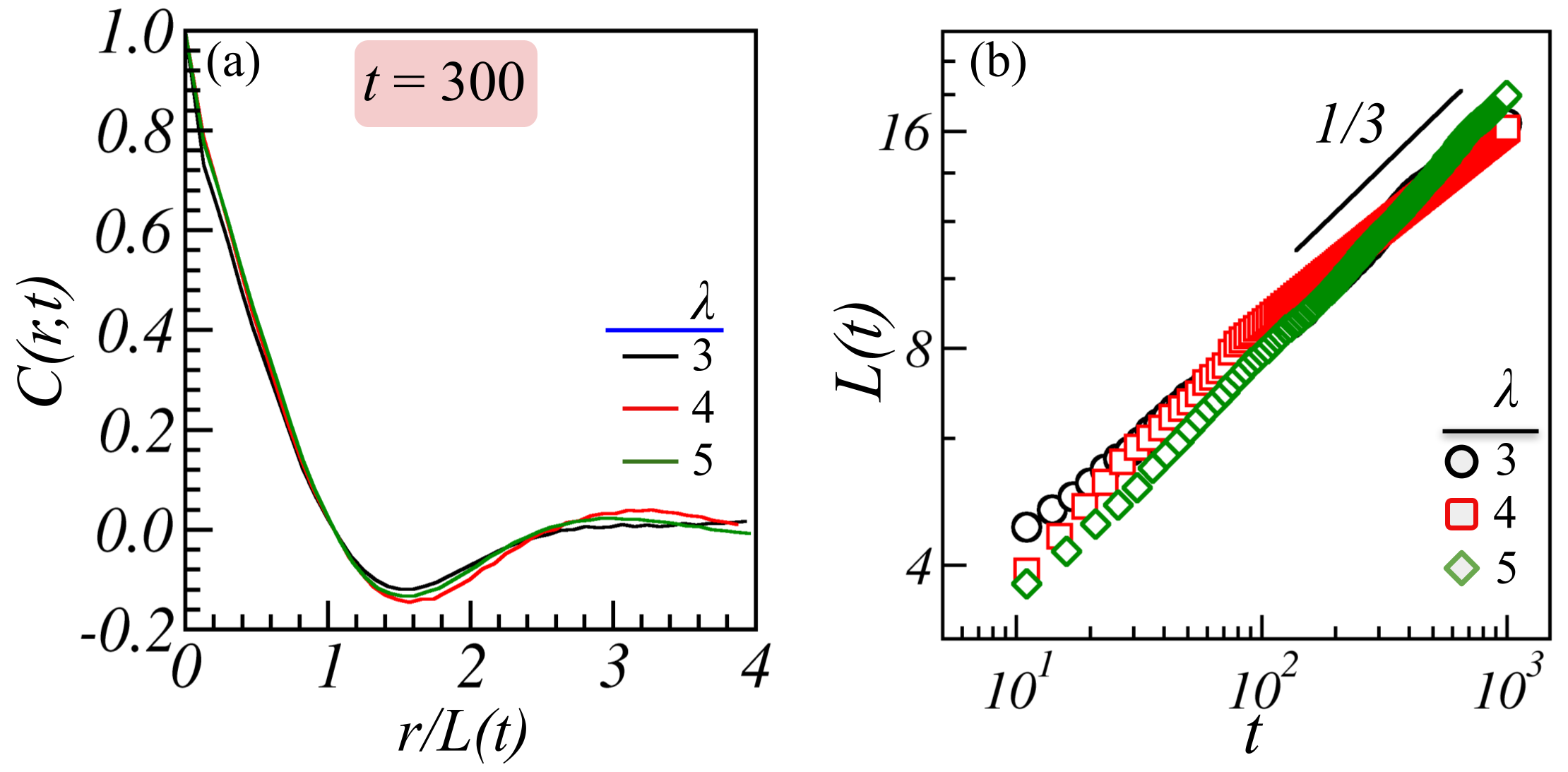}
\caption{(a) Scaling plots of the equal-time correlation functions $C(r,t)$ at $t=300$ for different values of $\lambda$, as mentioned. (b) Plot of average domain size $L(t)$ as a function of $t$ on a log-log scale for three different values of $\lambda$, as mentioned. The black solid line labeled $1/3$ represents the asymptotic $L(t)\sim t^{1/3}$ growth law.}
\label{Fig4}
\end{figure}

To further characterize the morphological transition observed in Fig.~\ref{Fig3}, we compute $C(r,t)$ for different values of $\lambda$. Figure~\ref{Fig4}(a) shows the scaled $C(r,t)$s at a fixed time $t = 300$ for the values of $\lambda$, as indicated. It is evident that the scaled master curves do not collapse onto a single universal function at large distances. This lack of data collapse indicates clear morphological differences in the domain structures as $\lambda$ is varied. The data mismatch at large $r$ translates to the small-$k$ regime of $S(k,t)$, while the tail of $S(k,t)$ follows Porod’s law, regardless of $\lambda$ (not shown here).

Next, we focus on how the average domain size $L(t)$ varies with time. Figure~\ref{Fig4}(b) shows $L(t)$ as a function of $t$ on a log–log scale for different values of $\lambda$, as indicated. After an initial transient regime, $L(t)$ grows as $L(t) \sim t^{1/3}$, independent of $\lambda$. In the context of phase separation in passive systems, this growth law is well known and has been observed during domain growth in off-critical binary mixtures, as well as during spinodal decomposition in critical mixtures~\cite{AO2002}. Recently, Mondal and Das reported the $L(t)\sim t^{1/3}$ growth law in the context of MIPS in active binary mixtures, irrespective of the relative activities of the species or their compositions~\cite{SMPD2024}.
\begin{figure}[tbh!]
\centering
\includegraphics[width=0.5\textwidth]{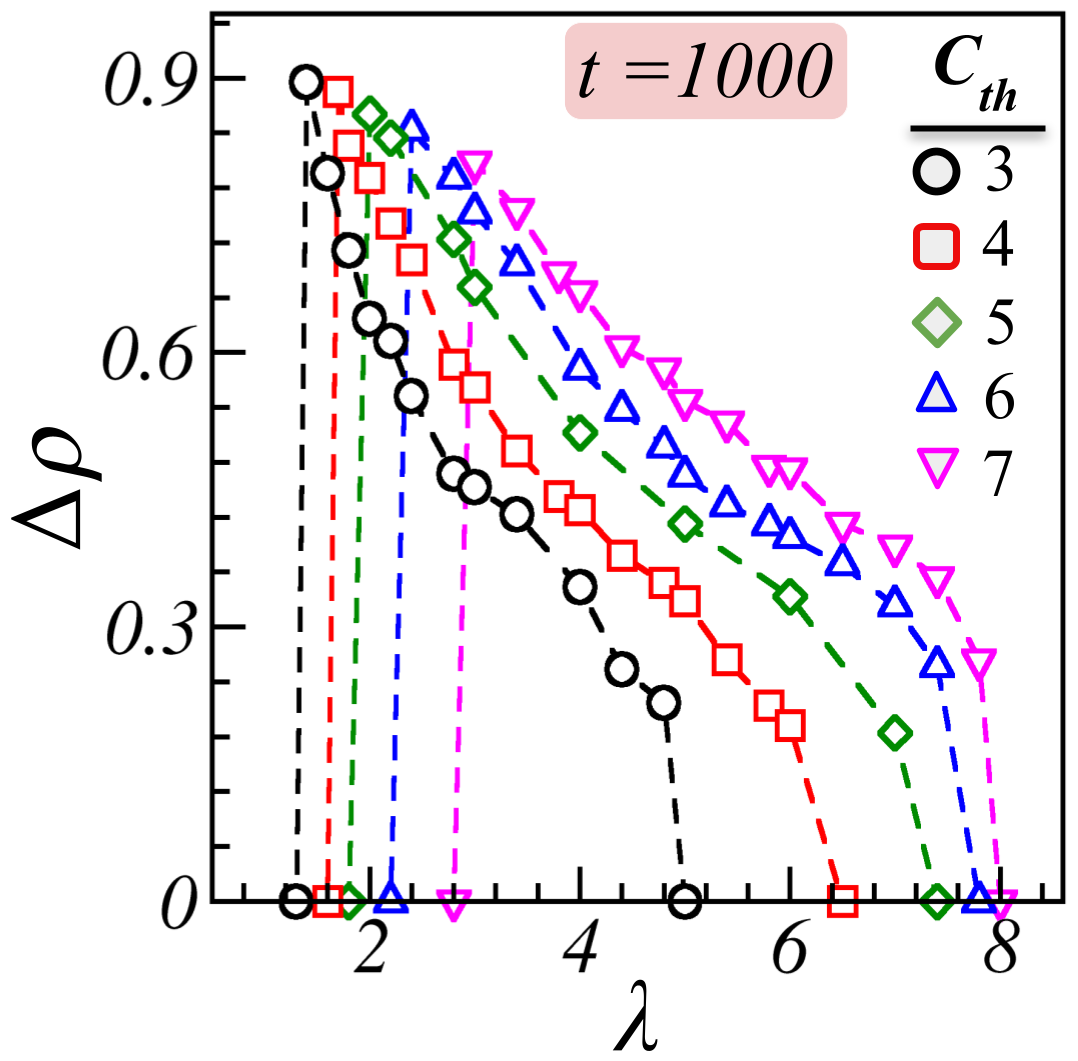}
\caption{Dependence of the average density difference, $\Delta \rho$, between the dense and dilute regimes of the system at a sufficiently late time ($t = 1000$) as a function of $\lambda$. The density of particles in the system is $\rho=0.2$.}
\label{Fig5}
\end{figure}

Further, we compute the density difference $\Delta \rho = \rho_1 - \rho_2$ as a function of $\lambda$ at a sufficiently late time. Here, $\rho_1$ and $\rho_2$ denote the particle densities in the dense and dilute regimes, respectively. We calculate $\rho_i$ as $\rho_i=\tfrac{n}{\pi s_c^2}$, where $n$ is the number of particles within a circle of radius $s_c$ chosen in the respective dense and dilute regimes. In simulation, we set $s_c=10\sigma$. The variation of $\Delta \rho$ with $\lambda$ enables us to determine the range of $\lambda$ values for which the system exhibits self-organization at a given $c_{\mathrm{th}}$. A value of $\Delta \rho \approx 0$ indicates that the system remains homogeneous, i.e., no phase separation occurs. Figure~\ref{Fig5} shows the variation of $\Delta \rho$ with $\lambda$ for different $c_{\mathrm{th}}$. Starting from $\lambda > r_{\rm cut}$, the system begins to exhibit phase separation. Clearly, $\Delta \rho$ decreases with increasing $\lambda$ for fixed $c_{\mathrm{th}}$. For sufficiently large $\lambda$ ($\gg r_{\rm cut}$), the quorum-sensing interaction sets the speed of all particles in the system to zero, leading to a frozen state for $c_{\mathrm{th}}$. Consequently, phase separation is no longer observed. The value of $\lambda$ at which the system freezes depends on $c_{\mathrm{th}}$. For very small-$\lambda$ ($<r_{\rm cut}$), the local concentrations $c_i(t)$ are unable to reach the quorum-sensing threshold $c_{\rm th}$, resulting in a state where all particles move with the propulsion speed $v_0$ in random directions, again preventing phase separation. Thus, the system exhibits phase separation only within a finite range of $\lambda$.

\begin{figure}
\centering
\includegraphics[width=0.90\textwidth]{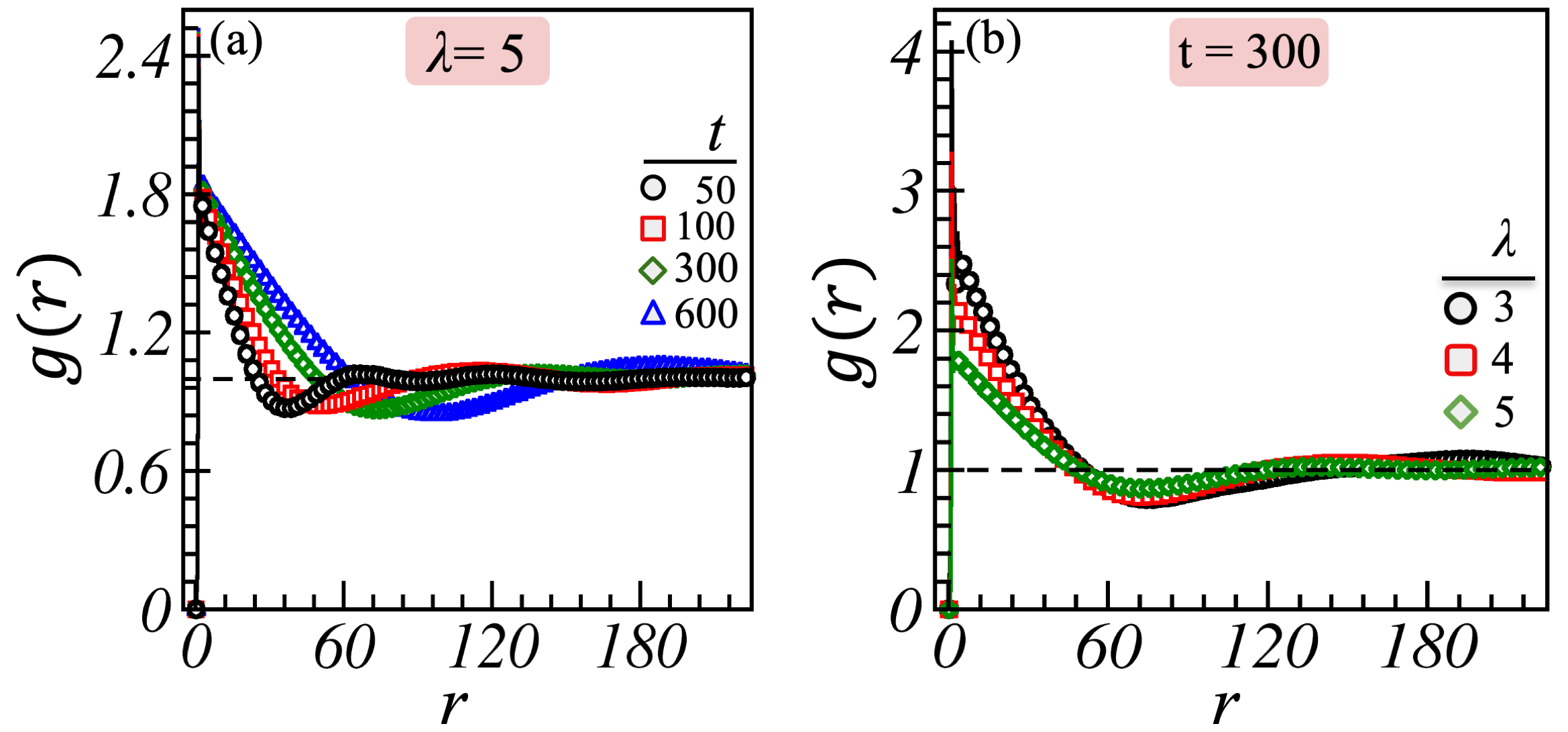}
\caption{Radial distribution function, $g(r)$, corresponding to (a) the evolution snapshots shown in Fig.~\ref{Fig2} at fixed $\lambda=5$ for different times, and (b) the snapshots shown in Fig.~\ref{Fig3} at fixed $t=300$ for $\lambda=3$, $4$, and $5$.}
\label{Fig3a}
\end{figure}

Finally, to characterize the dependence of $\Delta \rho$ on $\lambda$ in the self-organized phase, we compute the radial distribution function, $g(r)$, as a function of $r$, where $r$ is the radial distance. It describes the distribution of particles around a reference particle, and is defined as: $g(r)=\rho(r)/\tilde{\rho}$. Here, $\rho(r)$ represents the local density profile of particles at a distance $r$ around the reference particle, and $\tilde{\rho}$ is the average density of particles~\cite{McQuarrie,Avinash2023}. Figure~\ref{Fig3a}(a) shows the plot of $g(r)$ for the evolution snapshots shown in Fig.~\ref{Fig2}. At $r\approx \sigma$, all curves exhibit nearly the same peak height, suggesting that the domains at different times have a similar degree of packing. Furthermore, the decay of the curves slow down as the system evolves from the early time $t=50$ (black curve) to the later time $t=600$ (blue curve). This shift indicates an increase in the characteristic domain size as the domains grow over time. In Fig.~\ref{Fig3a}(b), we plot $g(r)$ versus $r$ corresponding to the snapshots shown in Fig.~\ref{Fig3}. At $r\approx \sigma$, the black curve for $\lambda=3$ the exhibits the highest peak in $g(r)$, whereas the green curve for $\lambda=5$ shows the lowest peak. The red curve corresponding to $\lambda=4$ exhibits an intermediate peak height. The difference in the peak heights of $g(r)$ suggests that the degree of domain packing decreases with increasing $\lambda$ from $3$ to $5$, indicating that $\Delta \rho$ decreases in the self-organized phase as $\lambda$ increases. In all cases, $g(r\rightarrow\infty)=1$, indicating a homogeneous density.

\section{Conclusions}\label{sec4}
Let us conclude this paper with a summary and discussion of our results. We have investigated the kinetics of self-organization in a one-component active matter system interacting via quorum-sensing interactions with $\lambda$ as the characteristic interaction length. In addition to quorum sensing, particles interact through excluded-volume interactions modeled by the WCA potential, and the direction of self-propulsion undergoes rotational diffusion. In this study, all model parameters are held fixed while only $\lambda$ and $c_{\mathrm{th}}$ are varied. We find that the system undergoes self-organization only within a finite range of $\lambda$ for a fixed $c_{\mathrm{th}}$. The system does not self-organize via quorum sensing when $\lambda<r_{\rm cut}$ or $\lambda\gg r_{\rm cut}$. Starting from random initial conditions, the system phase-separates into dense and dilute regions for appropriate values of $\lambda$.

At fixed $c_{\mathrm{th}}$, we observe a transition from circular to elongated domain structures as $\lambda$ increases. We quantify the asphericity of the domains using the eccentricity parameter, $e$. Small values of $e$ are associated with circular domains at low $\lambda$, while larger values of $e$ correspond to elongated or aspherical domain structures at higher $\lambda$. Intermediate values of $e$ characterize domain structures that lie between these two extremes, reflecting a gradual transition from droplet-like to elongated structures. Additionally, the compactness of the domains decreases with increasing $\lambda$, as confirmed by the radial distribution function $g(r)$. Next, we characterize the domain morphology by computing the equal-time density correlation function $C(r,t)$ and the corresponding structure factor $S(k,t)$, both of which exhibit dynamical scaling. However, the form of scaling functions depend on $\lambda$, indicating the occurrence of a morphological transition as $\lambda$ is varied. Furthermore, we calculate the average domain size $L(t)$ from the decay of $C(r,t)$. We find that $L(t)$ follows a power-law growth, $L(t)\sim t^{1/3}$, independent of $\lambda$. This observation suggests that the $t^{1/3}$ domain-growth law is robust with respect to the choice of physical model describing MIPS.

\ \\
\noindent{\bf Acknowledgments:} A. Chauhan is grateful to IISER-Mohali, India for the financial support through an institute postdoctoral fellowship.

\ \\
\noindent{\bf Conflict of interest:} The authors have no conflicts to disclose.

\end{document}